\documentclass[11pt,twoside]{article}

\usepackage{asp2014}

\aspSuppressVolSlug
\resetcounters

\begin{document}

\title{{\Large The Dark Side of the Sun} \\
A Plea for a Next-Generation Opacity Calculation}
\author{Regner Trampedach$^{1,2,3}$\\
\affil{$^1$Space Science Institute, 4750 Walnut St.,
          Boulder, CO 80301, USA; \email{rtrampedach@SpaceScience.org}}
\affil{$^2$Stellar Astrophysics Centre, Dept.\ of Physics and Astronomy,
          Ny Munkegade 120, Aarhus University, DK--8000 Aarhus C, Denmark}
\affil{$^3$Laboratory for Atmospheric and Space Physics, University of Colorado
          Boulder, 3665 Discovery Dr., Boulder, CO 80303, USA}}

% This section is for ADS Processing.  There must be one line per author.
\paperauthor{Regner~Trampedach}{rtrampedach@SpaceScience.org}{}{Space Science Institute}{}{Boulder}{CO}{80301}{USA}

%\aindex{Trampedach, Regner}

%%%%%%%%%%%%%%%%%%%%%%%%%%%%%%%%%%%%%%%%%%%%%%%%%%%%%%%%%%%%%%%%%%%%%%%%%%%%%%%
\begin{abstract}
  Is the Sun likely to have a more opaque interior than previously thought? The solar oxygen (or abundance) problem can be solved with higher interior opacities, reconciling abundance analyses based on 3D convective atmospheres with the helioseismic structure of the solar interior. This has been known for more than a decade, but last year we learned that the absorption by just iron may contribute 7\% more to the solar opacity at the bottom of the convection zone than predicted by any opacity calculation so far, and by OP05 in particular. I find that artificial changes to the absorption (calibrated against the iron experiment) by other elements in a solar mixture give an opacity increase of a shape and magnitude that can restore agreement between modern abundance analysis and helioseismology. This suggests that improved opacity calculations will solve the solar oxygen problem.
\end{abstract}

%%%%%%%%%%%%%%%%%%%%%%%%%%%%%%%%%%%%%%%%%%%%%%%%%%%%%%%%%%%%%%%%%%%%%%%%%%%%%%%
\section{Our Problem with the Sun and Attempts to Solve it}
%%%%%%%%%%%%%%%%%%%%%%%%%%%%%%%%%%%%%%%%%%%%%%%%%%%%%%%%%%%%%%%%%%%%%%%%%%%%%%%

The solar abundance determinations by \citet{AGS05}---updated by \citet[AGSS09,][]{AGSS2009} and refined by \citet{scott:SolarFeGroup, scott:SolarNa-Ca} and \citet{grevesse:AGSS09-HeavyElems}--- have disrupted the previous convergence of solar models towards observations \citep{bahcall:seism-AGS05, delahaye:SolarZ, guzik:MassLossLowZaccreteOvShoot}.
The main part of the previous convergence was between solar structure and evolution models and helioseismology \citep[e.g.,][]{GONG-Sci:sol-mod, schou:SunDiffRot, brun:HelioseismOvershoot,boothroyd:SSMuncerts,
di_mauro:SeismicTool}.

Solar \emph{atmosphere} modeling, on the other hand, has had persistent problems in that same time period, mostly hampered by being one dimensional (1D). The solar atmosphere is convective, and convection is inherently a three dimensional (3D) phenomenon \citep{trampedach:Rome2009} although classically treated in the \emph{mixing-length formulation} (MLT) by \citet{boehm:mlt}. This formulation adds at least three free parameters to the problem with the primary one, $\alpha$, typically being adjusted to reproduce the wings of the Balmer lines \citep{fuhrmann:Balmer-lines}. This gives values of $\alpha\approx 0.5$ and has little relation to the $\alpha\sim 1.8$ for a solar evolution calibration \citep{gough:MLT-calibr} or calibrations against 3D convection simulations \citep{ludwig:alfa-cal,trampedach:alfa-fit,magic:stagger-alpha}.

To generate synthetic spectra additional velocity broadening of spectral lines is needed to describe their shapes, introducing the free parameters of micro- and macro-turbulence \citep[e.g.,][]{gray:SpctrLineAnalysis}.  All these free parameters greatly reduce the predictive power of spectral synthesis from 1D atmosphere models, and also contribute a significant scatter in abundances derived from such synthesis \citep{jofre:SpectroscopicReliability}---a scatter that greatly exceeds the stated uncertainties.

The aforementioned abundance analyses, on the other hand, are performed on 3D hydrodynamic simulations \citep{nordlund:StaggerCode} of the convective, radiating solar atmosphere. This obviates the need for the free parameters required in 1D models as the Doppler broadening velocity-field arises naturally in the simulations  \citep{asplund:solar-Fe-shapes}. Indeed there are no free parameters available to fit the observed spectra. The simulations do depend on, for instance, numerical resolution \citep{asplund:num-res}, which is chosen as a balance between available computing resources and the need to resolve pertinent features in the simulations---the resolution would not, however, be decreased in order to fit a spectral line. The emergent spectral lines are asymmetric, as observed, and make it trivial to
spot even weak blends, e.g., the \ion{Ni}{i} blend with the forbidden [\ion{O}{i}] line at 630\,nm \citep{prieto:Solar[OI]-abund}. With the symmetric lines of 1D atmospheres, this blend was underappreciated as it could not be quantified. Agreement on the detailed line shapes also means that blends can be left out, and abundances can be determined from fits to the actual line profiles instead of the integrated quantity of equivalent widths that is blind to blends. Undetected blends, of course, result in overestimated abundances. \citet{grevesse:StopUsingGN93} provide a scathing critique of their own
classic solar abundances based on 1D atmosphere models.

%%%%%%%%%%%%%%%%%%%%%%%%%%%%%%%%%%%%%%%%%%%%%%%%%%%%%%%%%%%%%%%%%%%%%%%%%%%%%%%
\section{An Opacity Solution?}
%%%%%%%%%%%%%%%%%%%%%%%%%%%%%%%%%%%%%%%%%%%%%%%%%%%%%%%%%%%%%%%%%%%%%%%%%%%%%%%

\citet{bahcall:dCZ}, \citet{basu:SolarZ}, and \citet{montalban:AGS05SunMod} showed that a ${\sim}20$\% Gaussian opacity increase in the solar interior, peaking at the bottom of the convection zone, would restore the previous helioseismic agreement based on classic abundances. \citet{jcd:AsteroSeism} found the actual opacity-increase profile needed to restore helioseismic agreement with a AGSS09-mix solar model. Whether such opacity increases were at all realistic was not known at the time.

Measurements of the iron opacity under conditions similar to those at the bottom of the solar convection zone \citep{bailey:Fe150eVplus, bailey:ExtraFeOpac} gave values significantly higher than predicted by any theoretical calculation to date. By including their measured Fe absorption in a Rosseland average for a solar mixture with opacities by \citet[OP05,][]{OP05}, a 7\% increase over the theoretical tables is obtained.

%%%%%%%%%%%%%%%%%%%%%%%%%%%%%%%%%%%%%%%%%%%%%%%%%%%%%%%%%%%%%%%%%%%%%%%%%%%%%%%
\section{What if Other Elements Are also More Opaque?}
%%%%%%%%%%%%%%%%%%%%%%%%%%%%%%%%%%%%%%%%%%%%%%%%%%%%%%%%%%%%%%%%%%%%%%%%%%%%%%%

\begin{figure}% [htb]
  \centering
  \includegraphics[width=0.45\textwidth,clip=false]{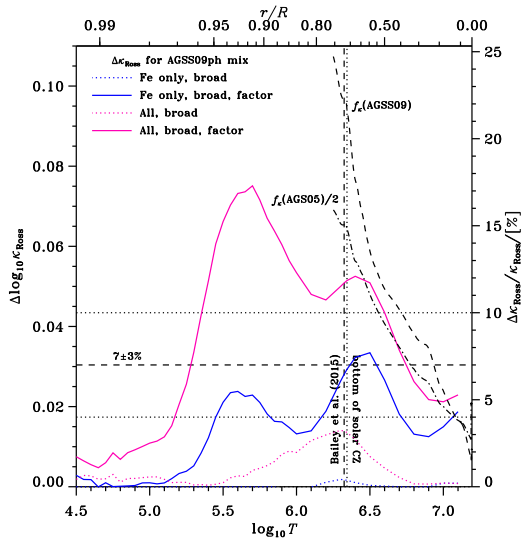}
  \caption{Opacity increase required to restore agreement with helioseismology (dashed and dot-dashed curves) along the
  stratification of a solar model. Blue dotted curve shows the opacity enhancement when only the absorption by iron is broadened to match the experiment \citep{bailey:ExtraFeOpac}. Solid blue curve when iron absorption is also increased by 15\% calibrated to match the experiment, indicated with the vertical dashed line. The magenta curves show the same but extended to all elements with the increase being proportional to the number of closed subshells.
  \label{opac_cmp_half_mini.withK}}
\end{figure}

We carried out an enhancement experiment on the absorption coefficients of the OP05 opacities: it was calibrated by broadening the OP05 absorption to qualitatively match the Fe absorption measured by \citet{bailey:ExtraFeOpac} and by multiplying the cross-sections by a constant factor to give the reported 7\% increase of the Rosseland mean. The crucial step is the extrapolation to the other elements and ions. Using the number of bound electrons in the radiating ion as a measure of our ignorance of atomic physics, the enhancement factor is made proportional to it (but excluding changes to the well-known hydrogen- and helium-like ions). For a solar stratification the thusly enhanced Rosseland mean is larger than the normal OP05 opacity by about 12\%, peaking just below the convection zone as shown in Fig.~\ref{opac_cmp_half_mini.withK}.
Both shape and amplitude are very similar to those needed to reconcile models with helioseismology \citep[][dot-dashed line in Fig.\ \ref{opac_cmp_half_mini.withK}]{jcd:AsteroSeism}. This result is not sensitive to how the bound electrons are counted, whether individually or in closed shells or subshells (as shown in Fig.~\ref{opac_cmp_half_mini.withK}), and simply reflects the abundance of bound electrons in a solar model. That the opacity increase required to solve the solar problem has this shape strengthens our suspicion that the opacity is indeed the culprit. If the required opacity change peaked instead in the core, it would be more natural to suspect problems with the nuclear reaction rates, their screening, or extra mixing processes.

Interestingly, the single dominant contributor to the opacity enhancement outlined above is iron, followed by nine other elements each peaking with about 1\% additional opacity spread across the solar radiative zone. This means the result is not contingent on every element behaving according to the assumptions of this numerical experiment. The underlying assumption of this numerical experiment is obviously the veracity of the \citet{bailey:ExtraFeOpac} experiment. We are eagerly awaiting confirmation from other experiments \citep[e.g., at NIF,][]{heeter:OpacAtNIF} and for other elements while appreciating the daunting tasks such experiments represent.

New calculations of Rosseland mean opacities have been published by the OPAS team \citep{lePennec:OPAS-SunMod} and the next-generation Los Alamos opacity team \citep{colgan:NextGenLosAlamosOpacs} resulting in better agreement with solar sound-speed profiles, but only by 13--25\% of that needed to reconcile models with seismology.
% New opacity calculations that describe the \citet{bailey:ExtraFeOpac}
% experiment, are also likely to solve the solar abundance problem, and have
% wide-ranging consequences for other stars,

\acknowledgements RT acknowledges funding from NASA grant NNX15AB24G. Funding for the Stellar Astrophysics Centre is provided by The Danish National Research Foundation (Grant DNRF106).

%\bibliography{bibs/conv,bibs/opac,bibs/eos,bibs/starmod,bibs/seism,bibs/obs,bibs/math,bibs/staratm}  % For BibTex

%\input{Trampedach.bbl}

\end{document}